# Efficient determination of the Hamiltonian and electronic properties using graph neural network with complete local coordinates

Mao Su[1,3], Ji-Hui Yang[1,2], Hong-Jun Xiang[1,2] and Xin-Gao Gong[1,2]

[1] Key Laboratory for Computational Physical Sciences (MOE), State Key Laboratory of Surface Physics, Department of Physics, Fudan University, Shanghai 200433, China

[2] Shanghai Qi Zhi Institute, Shanghai 200030, China

[3] Shanghai AI Laboratory, Shanghai 200030, China

**Abstract**: Despite the successes of machine learning methods in physical sciences, prediction of the Hamiltonian, and thus electronic properties, is still unsatisfactory. Here, based on graph neural network architecture, we present an extendable neural network model to determine the Hamiltonian from ab initio data, with only local atomic structures as inputs. Rotational equivariance of the Hamiltonian is achieved by our complete local coordinates. The local coordinates information, encoded using the convolutional neural network and designed to preserve Hermitian symmetry, is used to map hopping parameters onto local structures. We demonstrate the performance of our model using graphene and SiGe random alloys as examples. We show that our neural network model, although trained using small-size systems, can predict the Hamiltonian, as well as electronic properties such as band structures and densities of states (DOS) for large-size systems within the ab initio accuracy, justifying its extensibility. In combination with the high efficiency of our model, which takes only seconds to get the Hamiltonian of a 1728-atom system, present work provides a general framework to predict electronic properties efficiently and accurately, which provides new insights into computational physics and will accelerate the research for large-scale materials.

**INTRODUCTION**

The Hamiltonian lies at the heart of solid state physics as it determines the electronic properties, the eigenstates of a Hamiltonian can then be used to obtain other physical properties. For periodic systems, the Hamiltonian needs to be solved in reciprocal space, which then gives band structure information. The past half century has witnessed the development of solving Hamiltonian problems based on density functional theory (DFT)[1], which has achieved great success in understanding electronic properties. However, with increasing system size, the cost of solving Hamiltonian using DFT increases dramatically and has become one of the greatest challenges in solid state physics and materials science.

Machine learning can provide a possibility to solve such problems. However, despite past decades have witnessed many successes of machine learning methods in predicting physical properties[2-3] and interatomic potentials[4-6], prediction of the Hamiltonian, and thus electronic properties, is still unsatisfactory. Current models generally yield low accuracy and poor extensibility to predict electronic properties such as bandgaps, defects, and optical properties. Mapping the Hamiltonian onto some locality is crucial to design an extendable model[7-8]. Inspired by the tight binding method, the Hamiltonian can be represented by hopping parameters connecting the localized basis functions. Using numerical pseudo-atomic orbitals (PAOs) $\phi(\mathbf{r})$ as the basis[9], the hopping parameter between atoms $i$ and $j$ is defined by

$$h_{i\alpha,j\beta}^{(\mathbf{R}_n)} = \left\langle \phi_{i\alpha}\left(\mathbf{r} - \boldsymbol{\tau}_i\right) \middle| \hat{H} \middle| \phi_{j\beta}\left(\mathbf{r} - \boldsymbol{\tau}_j - \mathbf{R}_n\right) \right\rangle, \tag{1}$$

where $\alpha$ and $\beta$ denote the orbitals, $\tau$ is the atomic position and $\mathbf{R}_n$ is the lattice vector. Note that hopping parameters have locality and can be determined by local structures because hopping can be negligible when atomic distance is larger than a certain cutoff radius[7]. Therefore, the Hamiltonian matrix, which is constructed by hopping parameters, can be determined from local structures, and thus an extendable model for predicting the Hamiltonian is theoretically feasible. In previous studies, Hegde and Bowen proposed a model for predicting Hamiltonians in a small basis set and studied the s-orbital Cu and sp3-orbital diamond systems[10]. Schütt *et al.* developed the SchNOrb model for small, individual molecules with data augmentation[11]. Gu *et al.* introduced a neural network (NN) representation of tight-binding Hamiltonians for one-dimensional carbon chains[12]. Compared with the previous studies, we appropriately treat the equivariance, and thus significantly increase the prediction accuracy.

Because the Hamiltonian matrix is equivariant under rotations of coordinates, prediction of hopping parameters is more complicated than predicting measurable properties such as bandgaps, which are independent of the choice of coordinate system. Although the data augmentation technique can reflect rotations[11], it is obviously inefficient. To fix the rotation freedoms, Li *et al*. recently introduced the deepH model[13], in which the local coordinate (LC) for each atom pair are defined based on the local structures. The hopping parameters can be trained in the LC and then transformed back to the global coordinate to construct the Hamiltonian matrix. Using the atomic-like basis functions, the rotational transformations can be well described by the Wigner-D matrix. Nevertheless, more efficient methods for predicting the Hamiltonian are still being explored. For example, in the work of Li *et al*., hermiticity of the Hamiltonian is not properly considered. Moreover, the commonly defined LC is an incomplete descriptor, and this will be clarified in the following.

In this work, based on the graph neural network architecture and LC transformation, we have developed an extendable model, called graph LC-Net, to predict the hopping parameters and hence obtain DFT Hamiltonian. In addition to the rotational covariance, the key advantages of our model over similar works are as follows: the hermiticity of the Hamiltonian is utilized, a complete LC for each atom pair is defined by our new algorithm to avoid the degeneration problem for highly symmetric structures and the convolutional neural network (CNN) is used to handle the LC information. The node and edge features are updated by message passing with an attention mechanism. The output of the model is the triangular portion of the hopping parameter matrix, and then all the hopping parameters are assembled into the Hamiltonian matrix. Using graphene and SiGe random alloy systems as examples, we demonstrate that our graph LC-Net, trained using small-size systems, can predict the Hamiltonian as well as electronic properties such as band structures and densities of states (DOS) for large-size systems within the DFT accuracy, but with very little cost, i.e., it takes only about 10 seconds to get the Hamiltonian of a 1728-atom system using only one GPU card. In addition, the linear scaling with the number of atoms of graph LC-Net makes it very promising for studying electronic properties of large-scale systems.

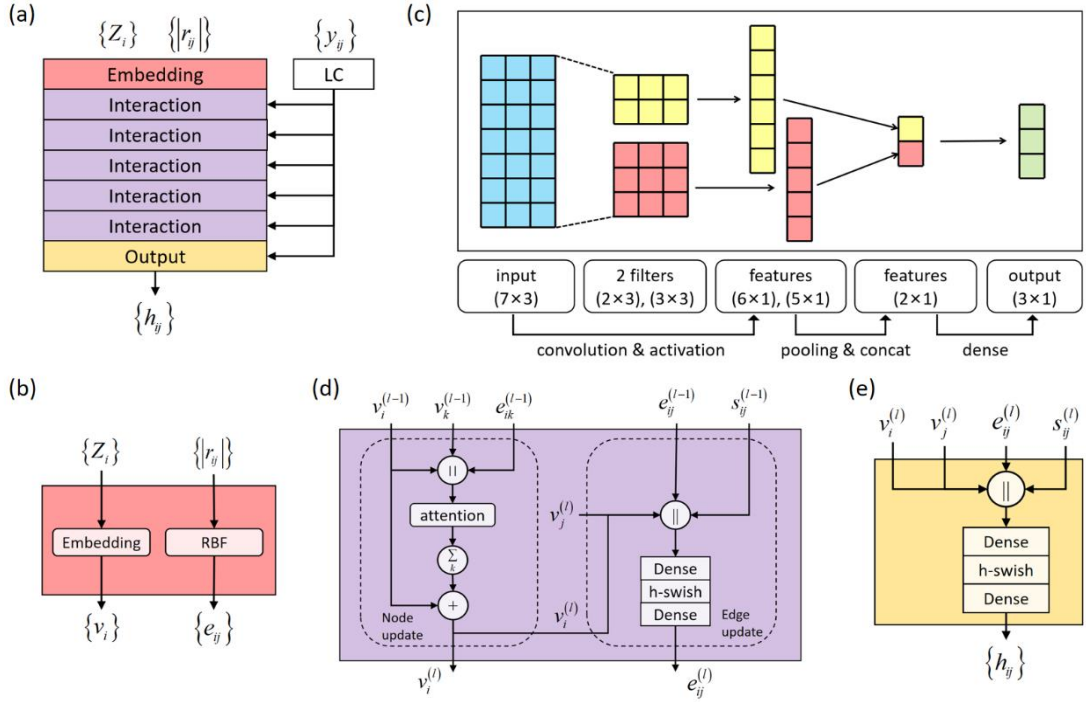

Figure 1. The architecture of graph LC-Net. (a) Graph LC-Net consists of an embedding layer, an LC layer, five interaction layers, and an output layer. The output of the LC layer is passed to each interaction layer. (b) Embedding layer. The atomic numbers are transformed to the initial node features through an embedding network. The interatomic distances are expanded by Gaussian radial basis functions to generate initial edge features. (c) An example LC layer with two filters. Spherical harmonics are used to generate LC features. (d) Interaction layer. The node features and edge features are updated using message passing. (e) Output layer. The node, edge and LC features are combined to output hopping parameters.

## RESULTS

**Embedding layer**

The atomic structure is represented by a graph, in which the nodes represent atoms, and the edges represent bonds between atoms. As shown in Fig. 1, the graph LC-Net model consists of an embedding layer, an LC layer, several interaction layers, and an output layer. The inputs include the atomic number $\{Z_i\}$, the edge distances $\{|r_{ij}|\}$, and the LC information $\{y_{ij}\}$. In the embedding layer, the initial node features $v_{ij}^{(0)} \in \mathbb{R}^{N_v}$ are calculated through an embedding network[14] by $v_i^{(0)} = a_{Z_i}$, and the initial edge features $e_{ij}^{(0)} \in \mathbb{R}^{N_e}$ are the Gaussian radial basis functions $e_{ij}^{(0)} = \exp\left(-\left(|r_{ij}| - r_n\right)^2 / \sigma^2\right)$, where the centers $r_n$ are placed linearly between 0 and the cutoff radius $R_c$, and $\sigma^2$ is the variance.

**Local coordinate (LC) layer**

The LC layer is used to obtain the LC information for each edge $e_{ij}$ which is uniquely determined by the local structures of atoms *i* and *j*. Note that, to utilize the hermiticity of the Hamiltonian, the edges *ij* and *ji* must be in the same LC. The details of calculating LC are described in the methods section. The LC information can be represented by real spherical harmonics and distances. Here we use the edge *ij* as an example to illustrate how LC information is obtained. First, the neighbors of atom *i*, denoted by *k*, are sorted by distances $|r_{ik}|$. Then we calculate the real spherical harmonics of bond *ik*, denoted by $Y_{lm}\left(\theta_{ik}^{ij}, \varphi_{ik}^{ij}\right)$, where $\theta_{ik}^{ij}$ and $\varphi_{ik}^{ij}$ are the polar and azimuthal angles, respectively, of bond *ik* relative to a LC defined for edge *ij*. Each bond *ik* can be represented by a list of real spherical harmonics, yielding the orientational vector $y_{ik}^{ij} \in \mathbb{R}^{N_i \times N_y}$ calculated by $y_{ik}^{ij} = Y_{lm}\left(\theta_{ik}^{ij}, \varphi_{ik}^{ij}\right) / (1 + |r_{ik}|)$, where $N_i$ is the number of neighbors of atom *i* and $N_y$ is the number of spherical harmonic functions. Similarly, for atom *j* we can obtain the orientational vector $y_{jk}^{ij} \in \mathbb{R}^{N_j \times N_y}$.

To extract the LC features $s_{ij}$ from the orientational vectors $y_{ik}^{ij}$ and $y_{jk}^{ij}$, Li *et al.* introduce the local coordinate message passing (LCMP) layer[13]. However, the order of the neighbors is not considered in the LCMP layer. To preserve the information of neighbor order, we utilize the CNN architecture for natural language processing (text-CNN)[15-16] as shown in figure 1(c). The convolution operator with a filter $w \in \mathbb{R}^{h \times N_y}$ is applied to a window of *h* neighbors to produce a feature. Using the convolution operator on the input $y_{ik}^{ij}$ generates a new feature map $c \in \mathbb{R}^{N_i - h + 1}$. Then max pooling is performed to extract a scalar from each feature map. These features are then passed to a dense layer to generate output vector $s_i \in \mathbb{R}^{N_{LC}}$. The neighbors of atom *j* are processed similarly, and finally the outputs are concatenated to give the LC feature vector $s_{ij} \in \mathbb{R}^{N_{LC} \times 2}$.

We use an example to further illustrate how to utilize the text-CNN to extract LC information.

Consider a local structure with respect to atom *i* in Fig. 2(a), the LC is defined on edge *ij*, and atoms 1 - 4 are neighbors of atom *i*. The input of the LC layer is the spherical harmonics $Y_{lm}^{i1}$, ..., $Y_{lm}^{i4}$. Each $Y_{lm}$ is a vector of $N_y = 25$ elements as $0 \leq l \leq 4$ and $-l \leq m \leq l$ are used. In analogy with the text-CNN framework, each neighbor is treated as a word in our model. Since the neighbors are sorted, we expect that the neural network is able to learn the information of order, which is not considered by the LCMP[13]. Nevertheless, the text-CNN layer could handle the serialized information appropriately. When a different LC is chosen, the input of spherical harmonics will be different, and thus the information of LC is learned. We have used three windows of sizes 2, 3 and 4, each of which has 64 filters. Thus, for each LC, 192 new features are extracted by the convolution operator. The output is a vector of $N_{LC}$ elements representing the LC information from the neighbors of atom *i*. The neighbors of atoms *j* are processed similarly. Finally, the outputs of atom *i* and atom *j* are concatenated to give LC feature $s_{ij}$ as shown in Fig. 2(b).

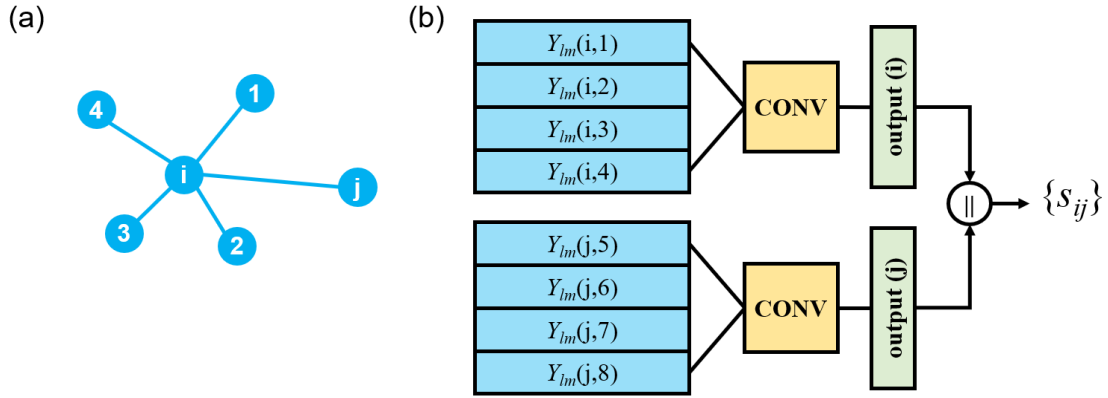

Figure 2. Illustration of using CNN to extract LC information. (a) The local structure of atom *i*. (b) Schematic of extracting the LC information of edge *ij*. The atoms 1 - 4 are neighbors of atom *i* and the atoms 5 - 8 are neighbors of atom *j*.

**Interaction layer**

After the processing of the embedding layer and the LC layer, information is passed to the multiple-stacked interaction layers. In the interaction layer, the attention mechanism[17-18] and skip connection are used for updating node features. Note that, our model is a universal framework, the interaction layers could be replaced by other advanced graph-based models for potentially better performance. To utilize the LC features, we just concatenate the LC feature $s_{ij}$ and the edge feature $e_{ij}$, since they are in one-to-one correspondence. As shown in figure 1(d), the node features $v_i \in \mathbb{R}^{N_v}$ are updated by

$$v_i^{(l)} = \sum_{k \in \mathcal{N}(i)} \alpha_{i,k} \Phi v_k^{(l-1)} + v_i^{(l-1)}, \tag{2}$$

where $k$ denotes the neighbor of $i$, $l$ refers to the $l$-th interaction layer, and $\Phi$ denotes a dense layer. The attention coefficients $\alpha_{i,k}$ are computed from

$$\alpha_{i,k} = \frac{\exp\left(\mathbf{a}^{\mathrm{T}} \mathrm{LeakyReLU}\left(\Phi[v_i \| v_k \| e_{ik} \| s_{ik}]\right)\right)}{\sum_{k \in \mathcal{N}(i)} \exp\left(\mathbf{a}^{\mathrm{T}} \mathrm{LeakyReLU}\left(\Phi[v_i \| v_k \| e_{ik} \| s_{ik}]\right)\right)}, \tag{3}$$

where $\mathbf{a}^{\mathrm{T}}$ is a learnable weight variable, LeakyReLU[19] is an activation function and $\|$ denotes concatenation. After the node features are updated, the edge features are updated from

$$e_{ij}^{(l)} = \Phi\left[\text{h-swish}\left[\Phi\left(v_i^{(l)} \| v_j^{(l)} \| e_{ij}^{(l-1)} \| s_{ij}\right)\right]\right], \tag{4}$$

where h-swish is the hard version of the swish activation function[20].

**Output layer**

The output layer consists of two stacked dense layers with h-swish,

$$h_{ij} = \Phi\left[\text{h-swish}\left[\Phi\left(v_i \| v_j \| e_{ij} \| s_{ij}\right)\right]\right], \tag{5}$$

where $h_{ij} \in \mathbb{R}^{N_h}$ is the upper triangular portion of the hopping matrix of $ij$. Since the Hamiltonian is Hermitian, the model does not need to output all matrix elements. However, the hopping matrix in general is not Hermitian, that is, $h_{i\alpha,j\beta} \neq h_{i\beta,j\alpha}^*$ unless $i = j$. Considering the hermiticity of the Hamiltonian, it can be shown that the hopping matrix satisfies $h_{i\alpha,j\beta} = h_{j\beta,i\alpha}^*$. Therefore, only the upper (or lower) triangular portion of hopping parameters are required as output of the model. There is a problem that, in general, the upper triangular portion of $h_{ij}$ and $h_{ji}$ are different, that is, $h_{ij} \neq h_{ji}$. The hermiticity can be utilized only if $h_{ij}$ and $h_{ji}$ are in the same LC. In this case, the input information for predicting $h_{ij}$ and $h_{ji}$ are the same, that is, $h_{ij}^{\text{Pred}} = h_{ji}^{\text{Pred}}$, and then the training will not converge. To overcome this problem, the hopping matrices are divided into two groups according to the distances of the neighbors, as described in the method section. The two groups are trained separately. Then we obtain two models for the hopping parameters, called *forward model* and *backward model*, respectively. Finally, the Hamiltonian matrix is constructed by all the hopping matrices, as shown in Fig. 3.

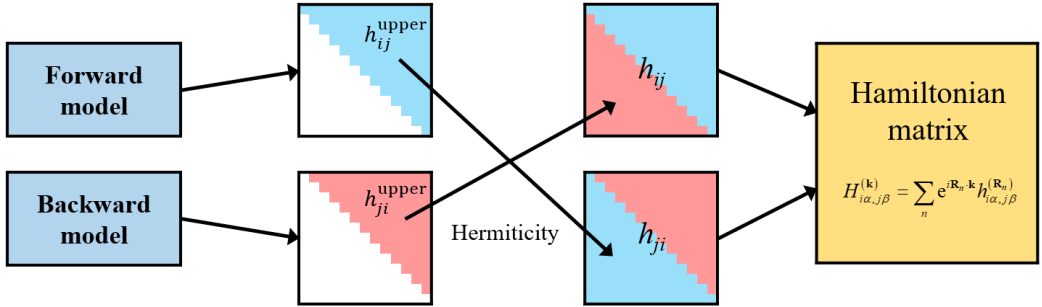

Figure 3. The hopping parameters are divided into two groups according to the definition of LC. We train a forward model for the group {$h_{ij}$} and a backward model for the group {$h_{ji}$} respectively. The two models share the same architectures and input information. The forward model outputs the upper triangular part with the main diagonal of {$h_{ij}$}, and the backward model outputs the upper triangular part without the main diagonal of {$h_{ji}$}. The lower triangular parts are determined by the hermiticity. Then all the hopping parameters are assembled into the Hamiltonian matrix.

**Loss function**

The graph CL-Net is trained by minimizing the difference between the predicted data $h_{ij}$ and the DFT calculated data $\hat{h}_{ij}$. We note that the hopping parameters are highly imbalanced, that is, most targets are close to zero while a few are larger than 10 eV. To better handle the imbalanced data, a regression version of the Focal loss[21-22] is used with

$$L(y,\hat{y}) = \frac{1}{n}\sum_{i=1}^{n}\left(\frac{1}{1+e^{-\beta|y_i-\hat{y}_i|}}\right)^{\gamma}|y_i-\hat{y}_i|, \qquad (6)$$

where $\beta$ and $\gamma$ are adjustable hyper-parameters, and $y$ denotes the output hopping parameters.

**Model evaluation**

With the above architecture for learning the Hamiltonian in hand, we turn to real systems to demonstrate the feasibility of learning electronic properties using the NN methods. The goal is to map the local atomic structures to the hopping parameters $h_{i\alpha,j\beta}^{(\mathbf{R}_n)}$ using graph CL-Net. In this work, s2p2d1 PAOs are used as basis functions and the hopping parameters for each atom pair $ij$ are represented by a 13×13 matrix. The mean absolute error (MAE) between DFT calculated and neural network hopping parameters is used to evaluate our method. The electronic properties including band structures and DOSs are then calculated for comparison.

**Degeneration problem with local symmetry**

The first step to train a Hamiltonian model is to define the LC for each atom pair. In the previous study[13], the LC is determined by the nearest neighbor $n_1$ and the second nearest neighbor $n_2$. However, this method only works on perturbated structures, that is, the nearest and second-nearest neighbors are unique. In the case of highly symmetric local structures, for instance, defect free crystals, the above method fails to determine a unique LC since the nearest neighbor cannot be determined uniquely by distances, and then results in data conflict. To be specific, for the neighbors of atom $i$, if $|\mathbf{r}_{n_1} - \mathbf{r}_i| = |\mathbf{r}_{n_2} - \mathbf{r}_i|$, the atoms $n_1$ and $n_2$ are indistinguishable. This is also known as the *degeneration problem*[23]. Similar problems can happen for the off-site hopping parameters. As a result, the errors of highly symmetric structures will be significantly larger, even in training set, than those of randomly perturbated structures due to data conflict. We find that this problem can be solved by utilizing the information of the global coordinate to define a *complete* LC for each edge, and the algorithm is provided in the method section.

To illustrate the feasibility of the new algorithm, we test the performance of different methods on a perfect graphene structure and a zincblende SiGe alloy. We first calculate LC for each edge using the original method of deepH[13], and then perform calculations with our new algorithm. The learning curves for graphene and SiGe are shown in Fig. 4(a) and Fig 4(b), respectively. It is shown that the training loss of the model with original LC is significantly larger than that with our complete LC.

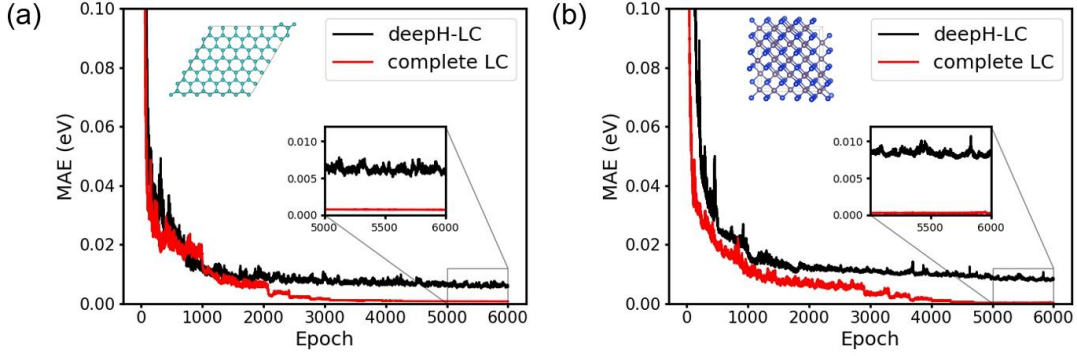

Figure 4. The learning curve of predicting the hopping parameters with different methods for LC. (a) The example of a perfect graphene structure. (b) The example of a zincblende SiGe alloy. Black line: the LCs are calculated using the method of deepH. Red line: the complete LCs are calculated using the algorithm proposed in this work.

**Graphene dataset**

We create a dataset of graphene to compare the performance of our graph LC-Net with the DeepH model proposed in ref.[13]. We perform molecular dynamics (MD) simulation of a 6×6×1 graphene structure for 5ps to generate dataset, and the computational details are consistent with the case of DeepH. Then we sample 500 structures as training set and other 500 structures as validation set. We find that, on the validation set, the MAEs of forward model and backward model are 3.5 meV and 1.9 meV, respectively, as shown in Fig. 5(a) and (b). Our result is comparable with that of DeepH (0.4 meV ~ 8.5 meV). The element-wise error distributions are shown in Fig. 5(c). The band structures and DOS are shown in Fig. 5(d). Note that, the authors of DeepH train 169 models for the hopping parameters to reconstruct the Hamiltonian of the graphene dataset, while we train only 2 models to obtain all the hopping parameters, demonstrating the efficiency of our method.

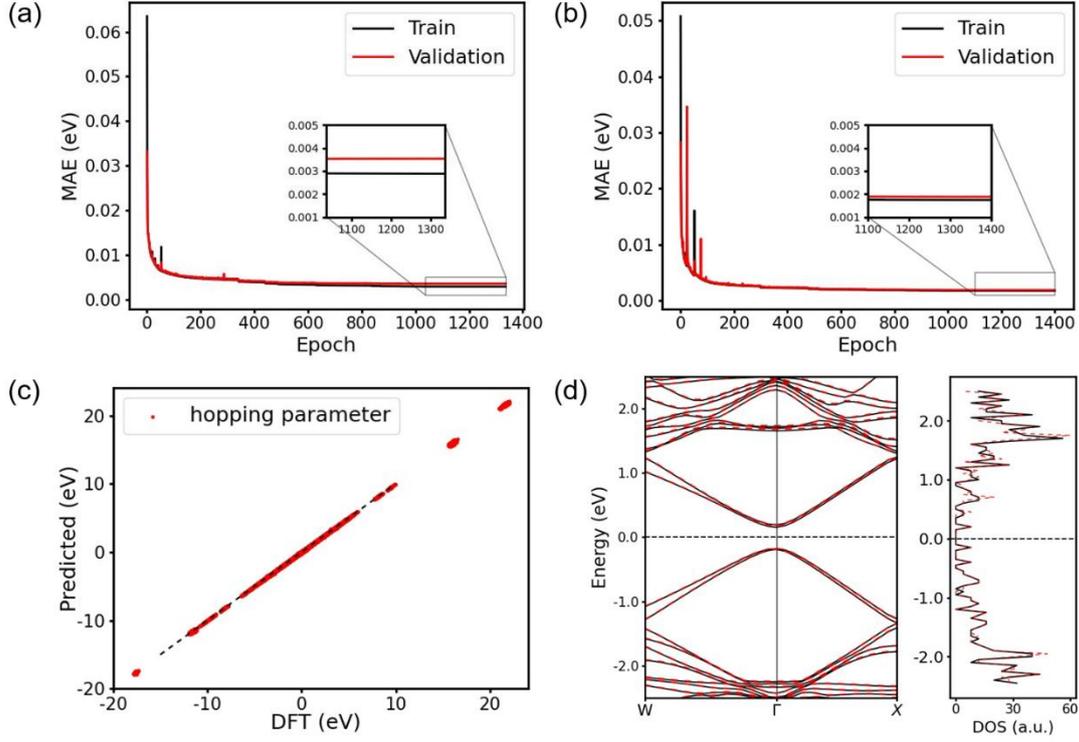

Figure 5. Model evaluation with the 6×6×1 graphene MD dataset. (a) The learning curve of forward model. The LCs are defined by equation (13) and (14). (b) The learning curve of backward model. The LCs are defined by equation (13) and (15). (c) Element-wise error distributions of the hopping parameters of a graphene in the validation set. (d) Band structure and DOS of a graphene in the validation set. Black line: results of graph LC-Net. Red dotted line: DFT results.

**SiGe random alloy dataset**

We demonstrate the feasibility of our method with the three-dimensional SiGe random alloy. The SiGe random alloy dataset are generated by randomly occupying the zinc-blende lattice sites with the Si or Ge atoms. The number of possible combinations in a supercell with $N$ sites is given by the combinatorial number $C(N, N/2)$, which could be incredibly large as the total atom number increases. Note that, most of the structures are identical under certain symmetry operations. Therefore, we develop a scheme to filter the identical structures to construct a training set of SiGe random alloy.

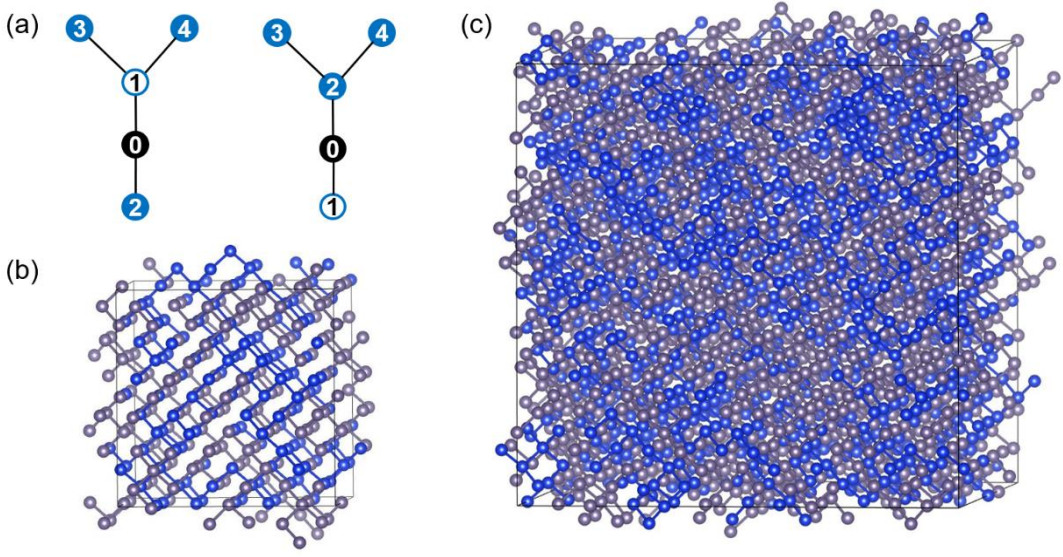

Figure 6. (a) Illustration of the two local structures of atom 0. The empty and full circles denote elemental type A and B, respectively. (b) Snapshot of a 3×3×3 SiGe random alloy in the training set. (c) Snapshot of a 6×6×6 SiGe random alloy in the validation set. All the structures are relaxed before calculating the Hamiltonians.

For each atom, we construct a unique local structure information by its neighbors in the following way. As shown in Fig. 6(a), we first calculate the dot product of $\mathbf{r}_{A,0}$ and $\mathbf{r}_{B,0}$ (atomic positions with respect to the target atom 0), and then obtain a list of dot products for the target atom:

$$v(0) = [\mathbf{r}_{10} \cdot \mathbf{r}_{20}, \mathbf{r}_{10} \cdot \mathbf{r}_{30}, \mathbf{r}_{10} \cdot \mathbf{r}_{40}]. \tag{7}$$

Then we sort the elements in the list $v(0)$ in an ascending order to represent the local structure information. It is easy to check that the two local structures in Fig. 6(a) are different by comparing the lists.

We calculate the local structure encodings by equation (7) for all atoms, remove duplicated ones and add the encodings into a database. When a new structure is generated, we compare the new encodings with those in the database. If new local structures are found, the generated structure is added into the training set and the database is updated accordingly. New structures are generated repeatedly until no more new local structures are found. In this work, we collect 66 samples of 2×2×2 supercells and 1000 samples of 3×3×3 supercells for training. A snapshot of a 3×3×3 SiGe random alloy in the training set are shown in Fig. 6(b). The samples in the validation set are randomly generated with supercell sizes ranging from 2×2×2 to 6×6×6. A snapshot of a 6×6×6 SiGe random alloy in the validation set are shown in Fig. 6(c). All the structures are relaxed before calculating the Hamiltonians. The structures in the training set contain up to 216 atoms (3×3×3 supercell). For the validation set, the structures contain up to 1728 atoms (6×6×6 supercell) and 25,406,784 hopping parameters, demonstrating the extensibility of graph CL-Net. For the training set, the MAEs of the forward model and the backward model are 0.50 meV and 0.52 meV, respectively. For the validation set, the MAEs range from 0.5 to 0.8 meV as shown in Fig. 7(a). The element-wise error distributions are shown in Fig. 7(b). The band structures and DOS are calculated using the Hamiltonian predicted by graph LC-Net, and comparisons with DFT calculations are given in Fig. 7(c) and (d).

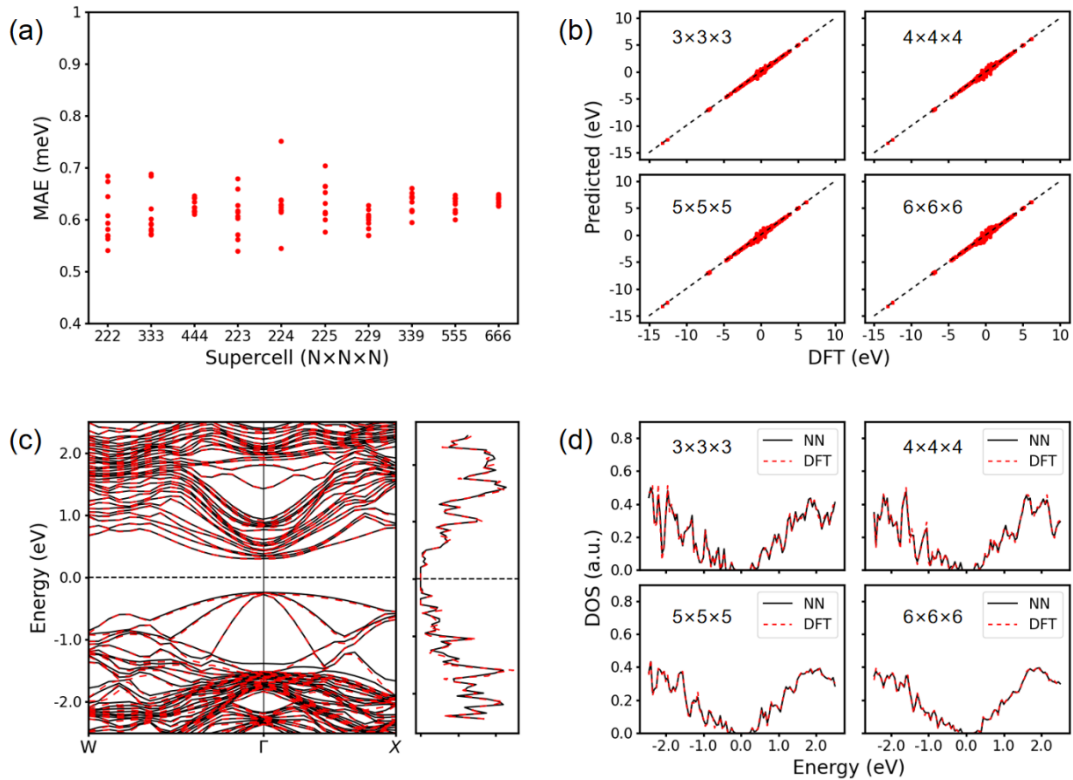

Figure 7. Model evaluation with the SiGe random alloy validation set. (a) The MAE results of hopping parameters. Each point corresponds to an average value of a supercell. (b) Element-wise error distributions of the hopping parameters of 3×3×3, 4×4×4, 5×5×5 and 6×6×6 supercells. Each point corresponds to an element of the hopping

matrix. (c) Band structure and DOS of a 2×2×2 supercell. (d) DOSs of 3×3×3, 4×4×4, 5×5×5 and 6×6×6 supercells. Black line: results of graph LC-Net. Red dotted line: DFT results.

## DISCUSSION

The major advantage of graph CL-Net over DFT calculation is the computational efficiency. The computational cost of graph CL-Net scales linearly with the number of atoms, compared with the cubic scaling of the DFT calculation. For the 6×6×6 supercell with 1728 atoms, graph CL-Net only costs 10.5 seconds to predict the Hamiltonian. For comparison, the computational time for the DFT self-consistent calculation is about 4800 seconds using 96 CPU cores, depending on the iteration steps. In this work, graph CL-Net is trained using one GeForce RTX 3090 card, and the DFT calculations are performed using the Intel Xeon Platinum 9242 CPU.

With DFT accuracy and very high computational efficiency, graph CL-Net can be applied to study electronic properties of very large systems, such as high-entropy alloys[24] or defect structures[25]. Graph CL-Net is also expected to accelerate inverse design of materials with targeted electronic properties[26-27] such as band gaps, where DFT calculations are performed thousands of times for structure relaxation and property evaluation. The structure relaxations can be efficiently performed using the NN potential models[28], and our graph CL-Net for Hamiltonian can be used to evaluate the electronic properties to avoid the computationally expensive DFT calculations.

In conclusion, based on the graph neural network architecture and using the local atomic structure information, we have developed the graph CL-Net to predict the hopping parameters and hence obtain DFT Hamiltonian. We have designed the complete LC algorithm for each edge to fix the rotational freedom of the Hamiltonian matrix and reserve the hermiticity of Hamiltonian to improve the efficiency. By employing graphene and SiGe random alloys as examples, we have demonstrated the high accuracy, extendibility, and efficiency of graph CL-Net in predicting Hamiltonian and electronic properties. This work thus opens doors for studying the electronic properties of large-scale systems using machine learning methods.

# METHODS

**DFT calculation**

Before calculating the hopping parameters, all the structures are relaxed into fixed cells using the Vienna *ab initio* Simulation Package[29]. The projector augmented wave (PAW)[30] type pseudopotential is used and the plane wave energy cut-off is set to 450 eV. The generalized gradient approximation (GGA) parameterized by Perdew, Berke and Ernzerhof (PBE)[31] and the local spin density approximation (LSDA) of Ceperley-Alder (CA)[32] are used for the exchange-correlation functional for graphene and SiGe alloy, respectively. The relaxation is stopped when the change of the total energy is smaller than 0.001 eV between two ionic steps.

Then we employ the numerical pseudo-atomic orbitals (PAOs) as implemented in OpenMX software[33] to perform DFT calculations for the hopping parameters. The PAO is given by a product of a radial function $R$ and a real spherical harmonic function $Y$ by

$$\phi(\mathbf{r}) = R(r)Y(\theta,\varphi), \tag{8}$$

where $R$ is defined numerically and is finite within a cutoff radius. The Hamiltonian and the overlap matrices are given by

$$\begin{aligned} H^{(\mathbf{k})}_{i\alpha,j\beta} &= \sum_n e^{i\mathbf{R}_n\cdot\mathbf{k}} \langle \phi_{i\alpha}(\mathbf{r}-\tau_i) | \hat{H} | \phi_{j\beta}(\mathbf{r}-\tau_j-\mathbf{R}_n) \rangle, \\ &= \sum_n e^{i\mathbf{R}_n\cdot\mathbf{k}} h^{(\mathbf{R}_n)}_{i\alpha,j\beta}, \end{aligned} \tag{9}$$

and

$$\begin{aligned} S^{(\mathbf{k})}_{i\alpha,j\beta} &= \sum_n e^{i\mathbf{R}_n\cdot\mathbf{k}} \langle \phi_{i\alpha}(\mathbf{r}-\tau_i) | \phi_{j\beta}(\mathbf{r}-\tau_j-\mathbf{R}_n) \rangle, \\ &= \sum_n e^{i\mathbf{R}_n\cdot\mathbf{k}} s^{(\mathbf{R}_n)}_{i\alpha,j\beta}, \end{aligned} \tag{10}$$

where $i$ and $\alpha$ denote atom and orbital, respectively, $\tau$ is the atomic position and $\mathbf{R}_n$ is the lattice vector. According to the principle of locality, the hopping parameters $h^{(\mathbf{R}_n)}_{i\alpha,j\beta}$ are determined by the local structures of atoms $i$ and $j$ within a certain cutoff radius. The overlap parameters $s_{i\alpha,j\beta}^{(\mathbf{R}_n)}$ are calculated by the PAOs without DFT calculations. Then the eigenvalues $v_{\mu\mathbf{k}}$ as well as eigenstates $E_{\mu\mathbf{k}}$ of a system are obtained by solving the generalized eigenvalue problem:

$$H^{(\mathbf{k})} v_{\mu\mathbf{k}} = E_{\mu\mathbf{k}} S^{(\mathbf{k})} v_{\mu\mathbf{k}}. \tag{11}$$

For graphene, the C6.0-s2p2d1 PAOs are used with cut-off radius of 6.0 Bohr as the basis functions. The Monkhorst–Pack k-meshes of 5 × 5 × 1 are used. For SiGe random alloy, the Si7.0-s2p2d1 and Ge7.0-s2p2d1 PAOs are used as the basis functions and the cut-off radius is set as 7.0 Bohr. For the supercells up to 4×4×4, the k-points are generated using the automatic scheme to ensure the spacing between k-points are smaller than 0.16 Å$^{-1}$. For the 5×5×5 and larger supercells, a single gamma point is used.

**Local coordinate (LC)**

First consider the case that the nearest neighbor and the second nearest neighbor are different atoms. We use two noncolinear vectors $\mathbf{v}_1$ and $\mathbf{v}_2$ to define the LC as

$$\hat{x}' = \frac{\mathbf{v}_1}{|\mathbf{v}_1|}, \hat{y}' = \frac{\mathbf{v}_1 \times \mathbf{v}_2}{|\mathbf{v}_1 \times \mathbf{v}_2|}, \hat{z}' = \hat{x}' \times \hat{y}'. \tag{12}$$

For onsite hopping ($i = j$), the two vectors are determined by

$$\mathbf{v}_1 = \mathbf{r}_{n_1} - \mathbf{r}_i, \mathbf{v}_2 = \mathbf{r}_{n_2} - \mathbf{r}_i, \tag{13}$$

where $n_1$ and $n_2$ denote the nearest neighbor and the second nearest neighbor of atom $i$, respectively.

For offsite hopping ($i \neq j$), we first calculate $d_{\min}(i)$, which is the distance from the nearest neighbor to the target atom $i$. To utilize the hermiticity, the LCs are defined in two different ways according to the distances. If $d_{\min}(i) < d_{\min}(j)$, we define

$$\mathbf{v}_1 = \mathbf{r}_j - \mathbf{r}_i, \mathbf{v}_2 = \mathbf{r}_{n_{i1}} - \mathbf{r}_i, \tag{14}$$

where $n_{i1}$ denotes the nearest neighbor of atom $i$. If $d_{\min}(i) > d_{\min}(j)$, we define

$$\mathbf{v}_1 = \mathbf{r}_i - \mathbf{r}_j, \mathbf{v}_2 = \mathbf{r}_{n_{j1}} - \mathbf{r}_j, \tag{15}$$

where $n_{j1}$ denotes the nearest neighbor of atom $j$. In this way, the $h_{ij}$ and $h_{ji}$ are ensured to be in the same LC. Since the input information for predicting $h_{ij}$ and $h_{ji}$ are the same, the hopping parameters defined by equation (14) and (15) are trained separately. The two models are called *forward model* and *backward model*, respectively.

As for the highly symmetric structures where degeneration problems happen, the nearest neighbor and the second nearest neighbor are the same atom, and therefore the above method is not applicable. In this case, we propose **Algorithm 1** with **Function find_v2()** to determine the basis vectors $\mathbf{v}_1$ and $\mathbf{v}_2$.

---

**Algorithm 1** Local coordinate
---
**Input:** Atomic positions $\{\mathbf{r}_i\}$, neighbors' distances $\{d_{ik}\}$ (sorted)
/* Find two vectors $\mathbf{v}_1$ and $\mathbf{v}_2$ to define the LC */
**if** $i = j$ **then**
  **if** $d_{i0} < d_{i1} < d_{i2}$ **then**    /* the nearest and second nearest neighbors are unique */
    The $\mathbf{v}_1$ and $\mathbf{v}_2$ are determined by the 1st and 2nd neighbors of atom $i$, respectively
  **else**
    Get candidates for $\mathbf{v}_1$: $\{\text{List } \mathbf{v}_1\}$ and $\mathbf{v}_2$: $\{\text{List } \mathbf{v}_2\}$
    **if** size of $\{\text{List } \mathbf{v}_1\} > 1$ **then**
      Find the vector with the largest x component as $\mathbf{v}_1$
      Compare y and z components in sequence if necessary
    **end if**
    **if** size of $\{\text{List } \mathbf{v}_2\} > 1$ **then**
      $\mathbf{v}_2 \leftarrow \text{find\_v2}(\{\text{List } \mathbf{v}_2\}, \mathbf{v}_1)$
    **end if**
  **end if**
**else**
  /* $i \neq j$, $\mathbf{v}_1$ could be either $\mathbf{r}_i - \mathbf{r}_j$ or $\mathbf{r}_j - \mathbf{r}_i$ */
  Calculate $\{\text{List } \mathbf{v}_2(i)\}$ and $\{\text{List } \mathbf{v}_2(j)\}$ [a]
  **if** size of $\{\text{List } \mathbf{v}_2(i)\} = 1$ **then**
    $\mathbf{v}_1 \leftarrow \mathbf{r}_j - \mathbf{r}_i$
  **else if** size of $\{\text{List } \mathbf{v}_2(j)\} = 1$ **then**
    $\mathbf{v}_1 \leftarrow \mathbf{r}_i - \mathbf{r}_j$
  **else**
    $\mathbf{v}_2(i) \leftarrow \text{find\_v2}(\{\text{List } \mathbf{v}_2(i)\}, \mathbf{r}_j - \mathbf{r}_i)$
    $\mathbf{v}_2(j) \leftarrow \text{find\_v2}(\{\text{List } \mathbf{v}_2(j)\}, \mathbf{r}_i - \mathbf{r}_j)$
    $\mathbf{v}_2 \leftarrow \mathbf{v}_2(i)$ or $\mathbf{v}_2(j)$ [b]
  **end if**
---
[a] $\{\text{List } \mathbf{v}_2(j)\}$: candidates for $\mathbf{v}_2$ if $\mathbf{v}_1 = \mathbf{r}_i - \mathbf{r}_j$
[b] The function **find_v2()** calculates the angle between $\mathbf{v}_2$ and an auxiliary vector $\mathbf{v}_a$, chose the one with smaller angle

| **Function find_v2({List $v_2$}, $v_1$)** |
|---|

**Inputs:**
  {List $v_2$} : candidates for $v_2$
  $v_a$: an auxiliary vector[a]
**Procedure:**
  Calculate the rotation matrix **R** to transform $v_1$ to (1, 0, 0)
  $\alpha \leftarrow \pi$
  **For v** in {List $v_2$} **do**
    Calculate the vector **v** in rotated coordinate
    Calculate the angle $\alpha$ between **v** and $v_a$
    **if** a smaller $\alpha$ is found **then**
      $v_2 \leftarrow v$
    **end if**
  **end for**

[a] The auxiliary vector should be chosen to avoid ambiguities, i.e., there are not two or more vectors in the $v_2$ list share the same $\alpha$. In this work, $v_a$ = (3.4, 4.5, 5.6) is used.

**Rotation of Hamiltonian**

The rotation transformation with real spherical harmonics basis is rather complicated.[34] To be convenient, we first convert real spherical harmonics basis into complex spherical harmonics basis, and then the rotation transformation $R$ can be done with the Wigner D-matrix $D^l_{mm'}(R)$, where $l$ is the angular quantum number and $m$ is the magnetic quantum number. The rotation of a complex spherical harmonics is given by

$$Y_l^m(\mathbf{r}') = \sum_{m'=-l}^{l} \left[ D^l_{mm'}(R) \right]^* Y_l^{m'}(\mathbf{r}). \tag{16}$$

Note that the Wigner D-matrix is unitary, that is, $D_{mm'}(R)^* = D_{m'm}(R^{-1})$. The rotation transformation of the hopping parameter matrix $h_{i\alpha, j\beta} \equiv \langle \phi_{i\alpha} | \hat{H} | \phi_{j\beta} \rangle$ is given by

$$h'_{i\alpha, j\beta} = \sum_{a,b} D^{l_\alpha}_{\alpha,a}(R^{ij}) h_{ia, jb} D^{l_\beta}_{b,\beta}\left((R^{ij})^{-1}\right), \tag{17}$$

where $i, j$ denote atoms, and $a, b, \alpha, \beta$ denote orbitals. Finally, the hopping parameters are converted back to real spherical harmonics basis.

**Training details**

The graph CL-Net model is developed using the Pytorch-Geometric[35] Python library. The variance of Gaussian radial basis function $\sigma^2$ is set to 0.0044 Å$^{-2}$. The parameters $\beta = 0.25$ and $\gamma = 2.0$ are used in the loss function. Adam optimizer[36] was used with initial learning rate of 0.001. The reduce on plateau strategy is used to schedule the learning rate, and the lower bound of the learning rate is $1\times10^{-5}$. The number of trainable parameters is 524731.

**Conflict of interest**

The authors declare no conflict of interest.

**Acknowledgements**

This work was supported in part by the National Natural Science Foundation of China (Grant No. 12188101, 61904035, 11974078, 11991061). Computations are performed at the Supercomputer Center of Fudan University.

**Author contributions.**

M.S., J.Y. and X.G. conceived the idea and designed the research. M.S. performed the DFT calculations and implemented the codes. X. G. supervised the work. All authors discussed the results and were involved in the writing of the manuscript.

**Data availability**

The code and data necessary to reproduce the results in this work will be available online after the manuscript is accepted.